\documentclass[aoas,MSNbibl,nameyear,dvips]{arximspdf}

\doi{10.1214/12-AOAS558} 
\volume{6}
\issue{2}
\pubyear{2012}
\firstpage{429}
\lastpage{431}

\begin{document}
\begin{frontmatter}

\title{Special section: Statistical methods for next-generation gene sequencing data}
\runtitle{Special section}

\begin{aug}
\author[A]{\fnms{Karen} \snm{Kafadar}\corref{}\ead[label=e1]{kkafadar@indiana.edu}}
\runauthor{K. Kafadar}
\affiliation{Indiana University}
\address[A]{Department of Statistics \\
Indiana University \\
Bloomington, Indiana 47408-3825\\
USA\\
\printead{e1}} 
\end{aug}

\received{\smonth{4} \syear{2012}}


\end{frontmatter}

This issue includes six articles that develop and apply
statistical methods for the analysis of gene sequencing data
of different types.  The methods are tailored to the different
data types and, in each case, lead to biological insights not
readily identified without the use of statistical methods.
A~common feature in all articles is the development of
methods for analyzing simultaneously data of different types
(e.g., genotype, phenotype, pedigree, etc.); that is, using
data of one type to inform the analysis of data from
another type.

In the first article of this section, Li et al. address
the problem of multiple missing genotype data through a
Bayesian hierarchical approach. The goal is to
impute missing values in association studies between
genotypes (as measured by single nucleotide polymorphisms,
or SNPs, in DNA sequences) and phenotypes. Because missing
SNP information is common, case-wise deletion is, at best,
impractical, and often wasteful of valuable information
when SNP information is available for the rest of the case.
Li et al.  develop a~computationally-efficient approach to
multiple imputation of many missing SNPs that uses all
available phenotype information. They show that their
Bayesian Association with Missing Data (BAMD) approach
achieves the desired goal in that it enables efficient
detection of SNPs that are highly associated with phenotypes.

Zhou and Whittemore propose likelihood-based methods to
improve the accuracy of genotype calls using information
on linkage disequilibrium (LD) and Mendelian pedigree
information, particularly for multiple SNPs that exhibit
high LD (SNPs for which the squared correlation coefficient
between them is close to~1).  Thus, use of LD or pedigree
information can modulate the negative effects of errors in
sequence reads and alignments and hence enable better inference.
The approach is applied to data from both simulations and the
``1000 Genomes Project'' and is shown to improve the estimates
of model parameters and hence the accuracy of genotype calling.

Shen and Zhang develop a change-point model based on a nonhomogeneous
Poisson process (NHPP) to model sequence-read data on
DNA copy number variants (CNVs) in normal
(reference) samples, versus copy number aberrations (CNAs) in treated
(target) samples. Regions of intensity shifts in the NHPP may signal
regions of genetic polymorphisms that may be related to the differences
between target and reference samples.  Using a generalized likelihood
ratio statistic and a modified Bayes information criterion to select
the appropriate number of change-points, Shen and Zhang construct
Bayesian point-wise ``credible intervals'' to assess quantitatively the
effects of meaningful copy number estimates.  The high-throughput
nature of sequencing data necessitates computationally efficient
algorithms which are applied to sequencing data from tumor and normal
cell lines.

Zhou et al. also develop an approach to the analysis of
multiple gene expression studies which were conducted to
identify differentially expressed (DE) genes. In their
article, the authors use Bayesian model averaging (BMA) with
empirically-based prior model probabilities; simulations
demonstrate improved performance (sensitivity, specificity)
of DE gene detection using BMA versus one-at-a-time
single-model approaches. Applied to two microarray data sets,
the results identify DE genes related to lung disease with
covariates (smoking status, gender, race)
than found by either data set alone.

With multiple lists of apparently ``significantly'' differentially
expressed genes, Natarajan et al. describe approaches to quantify
the ``significance'' of top-ranked genes that appear
to be influential in several studies. Given $N$ studies,
in each of which the effects of $T$ genes are studied
and ranked, in how many of those $N$ studies would we
expect $n$ of the $r$ top-ranked genes to appear?
Natarajan et al. use the Poisson distribution, expected
sensitivity, and false discovery rate (FDR) to
characterize the significance of the size of the set
of frequently-appearing genes in the $N$ studies and
illustrate their inference method on studies on prostate
cancer gene expression.

In the final article of this special section,
%
Telesca et al. take a Bayesian approach to characterize
dependence among genes, and use directed graphs to account
for this dependence to explore, and develop inferences about,
differences in dependencies among genes. The
approach is motivated by a~search for genes related to
specific pathways that have been identified in the
progression of ovarian cancer.

High-throughput or next generation sequencing data,
with their high levels of sample dependence, size, and
dimension, present challenges to conventional statistical
methods that typically assume independence and ``$n > p$''
(more samples than dimensions). The features of these
complex data sets stimulate the development of methods
that can be used for model estimation, inference,
and the identification of biological mechanisms.\
Future issues will include additional articles on
novel methodology to address these and other challenges
posed by high-dimensional sequencing data.

\printaddresses

\end{document}